\documentclass[aip,amsmath,amssymb,reprint]{revtex4-1}

\usepackage[dvips]{graphicx}
\usepackage{dcolumn}
\usepackage[utf8]{inputenc}
\usepackage[T1]{fontenc}
\usepackage{mathptmx}
\usepackage{etoolbox}
\usepackage{siunitx}
\usepackage{color}
\usepackage{upgreek}
\usepackage{xspace}

\newenvironment{fequation*}{\empheq[box=\fbox]{equation*}}{\endempheq}
\newenvironment{falign*}{\empheq[box=\fbox]{align*}}{\endempheq}
\newenvironment{fmultline*}{\empheq[box=\fbox]{multline*}}{\endempheq}
\renewcommand{\bm}[1]{\boldsymbol{\mathbf{#1}}} \newcommand{\ud}{\mathrm{d}}
\newcommand{\bra}{\left\langle}
\newcommand{\ket}{\right\rangle}
\newcommand{\tens}[1]{\mathop{\buildrel \leftrightarrow \over{\bm{#1}}}\nolimits}
\renewcommand{\Re}{\operatorname{Re}} 
\DeclareMathOperator{\sinc}{sinc}

\newcommand{\ie}{i.e.\@\xspace}

\usepackage[normalem]{ulem}

\makeatletter
\def\@email#1#2{%
 \endgroup
 \patchcmd{\titleblock@produce}
  {\frontmatter@RRAPformat}
  {\frontmatter@RRAPformat{\produce@RRAP{*#1\href{mailto:#2}{#2}}}\frontmatter@RRAPformat} {}{}
  }%
\makeatother

\begin{document}

   \preprint{AIP/123-QED}

   \title{A diffusion model for light scattering in ejecta}

   \author{J.A. Don Jayamanne}
   \affiliation{CEA DIF, Bruyères-le-Châtel, 91297 Arpajon Cedex, France}
   \affiliation{Institut Langevin, ESPCI Paris, PSL University, CNRS, 75005 Paris, France}
   \author{J.-R. Burie}
   \email{jean-rene.burie@cea.fr}
   \author{O. Durand}
   \affiliation{CEA DIF, Bruyères-le-Châtel, 91297 Arpajon Cedex, France}

   \author{R. Pierrat}
   \affiliation{Institut Langevin, ESPCI Paris, PSL University, CNRS, 75005 Paris, France}
   \author{R. Carminati}
   \email{remi.carminati@espci.psl.eu}
   \affiliation{Institut Langevin, ESPCI Paris, PSL University, CNRS, 75005 Paris, France}
   \affiliation{Institut d'Optique Graduate School, Paris-Saclay University, 91127 Palaiseau, France}

   \date{\today}

   \begin{abstract}
      We derive a diffusion equation for light scattering from ejecta produced by extreme shocks
      on metallic samples. This model is easier to handle than a more conventional model based
      on the Radiative Transfer Equation (RTE), and is a relevant tool to analyze spectrograms
      obtained from Photon Doppler Velocimetry (PDV) measurements in the deep multiple
      scattering regime. We also determine the limits of validity of the diffusive model
      compared to the RTE, based on a detailed analysis of various ejecta properties in
      configurations with increasing complexity.
   \end{abstract}

   \maketitle

   \section{Introduction}\label{introduction}
     
\begin{figure}[!htb]
     \centering
     \includegraphics[width=0.6\linewidth]{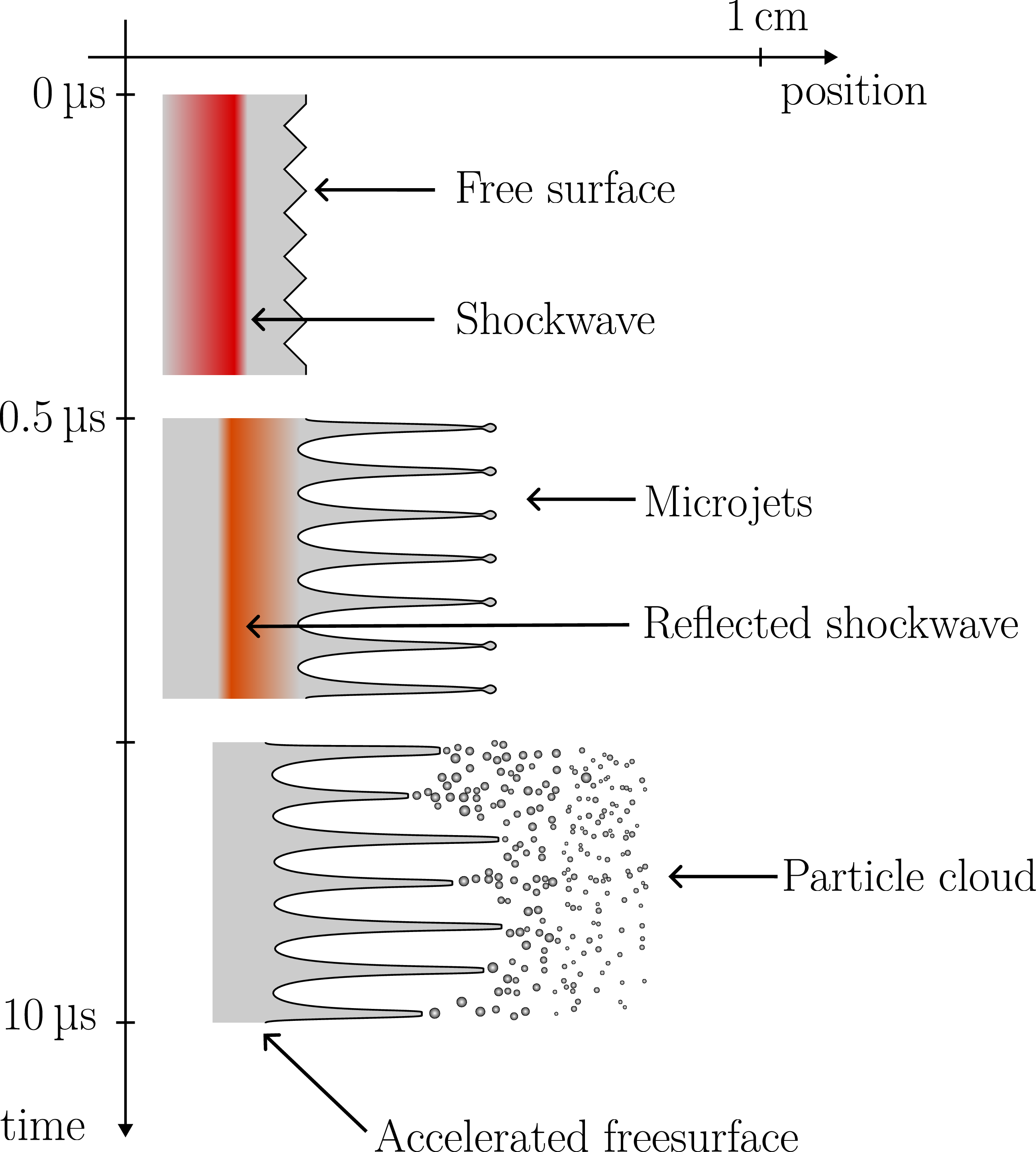}
     \caption{Illustration of the microjetting mechanism in a typical shock ejecta experiment.
     Upon reaching the machined free surface, the shock wave first comes into contact with the
     inwardly directed grooves. Under right angle conditions, the shock wave is reflected and
     the inward grooves become outward microjets.  Due to the velocity gap between the jet-heads
     and the free surface, the microjets are stretched until surface tension is no longer
     sufficient to hold matter together and fragmentation begins. This results in the creation
     of a particle cloud, \ie, an ejecta.}
     \label{ejection}
   \end{figure}

   Ejection is a naturally occurring phenomenon when a metallic sample is subjected to an
   extreme shock. Through an explosive~\cite{zellner_influence_2009} or high intensity
   laser~\cite{de_resseguier_microjetting_2014} solicitation, a shockwave can be released on one
   side of the sample with a typical shock pressure $P_s=\SI{10}{}-\SI{100}{\giga Pa}$.  Upon
   reaching the other side, called the free surface, the shockwave's interaction with the
   surface irregularities causes matter to partially melt, creating numerous expanding
   microjets. These microjets eventually fragment giving birth to a particle cloud called an
   ejecta~\cite{asay_ejection_1976,andriot_ejection_1982}. A schematic representation of this
   microjetting mechanism is depicted from top to bottom in Fig.~\ref{ejection}. Ejection has
   been extensively studied for the last fifty years~\cite{buttler_foreword_2017} and is now
   understood as a limiting case of Richtmyer-Meshkov
   instabilities~\cite{richtmyer_taylor_1960,meshkov_instability_1972}. One of the current goal
   of ejecta study is to better characterize the particle cloud through its number density, size
   or velocity distributions.

   Photon Doppler Velocimetry (PDV) is one of the optical diagnostics used in this
   characterization effort. Initially developed to monitor particle
   velocities~\cite{strand_compact_2006,mercier_photonic_2006} in the single scattering regime,
   the PDV spectrogram of an experiment can be seen as a time-velocity cartography of the
   ejecta. Recently, we have shown that PDV spectrograms can also be interpreted in the multiple
   scattering regime even if they do not give direct access to the particle velocity
   distribution~\cite{don_jayamanne_characterization_2024, don_jayamanne_recovering_2024}. In
   this context and from a theoretical point of view, PDV spectrograms are solutions of a light
   transport model valid from the single scattering to the diffusive regime, and based on the
   Radiative Transfer Equation (RTE)~\cite{CHANDRASEKHAR-1950,carminati_principles_2021}.

   Although the RTE is a highly relevant model for describing light propagation in a wide range
   of transport regimes, it remains difficult to manipulate. In practice, ejecta often have
   large optical thicknesses (typically on the order of $40$). In such thick samples, a
   simplified model derived from the RTE and called the diffusion
   approximation~\cite{ishimaru_wave_1997,carminati_principles_2021} is commonly used to
   describe the propagation of light. This model is much simpler to handle, and in some cases
   even admits of analytical solutions. The purpose of this work consists in deriving a
   generalized diffusion equation valid for ejecta. The main idea is to adapt the standard
   derivation of the diffusion equation from the RTE by taking into account the specificities of
   ejecta (in particular the statistical inhomogeneities and dynamic nature of the particle
   cloud). We will show how the diffusion approximation allows us to describe the PDV
   spectrograms much more easily and test the limits of this simplified model.

   The paper is organized as follows. Section~\ref{pdv} is dedicated to the introduction of PDV,
   the associated spectrograms and the theoretical model based on the RTE, valid from the single
   scattering to the deep multiple scattering regimes. In Sec.~\ref{diffusion}, we perform the
   diffusion approximation to obtain a diffusion equation for light propagation in ejecta. We
   compare the result to the standard diffusion equation and see especially how the displacement
   of the scatterers is effectively accounted for. Finally, a comprehensive study comparing the
   results given by the RTE and the diffusion equation for different dynamic media is reported
   in Sec.~\ref{results}. To check the robustness of the diffusion model, we start from the
   simplest configuration and progressively increase the complexity to eventually treat the case
   of a real ejecta.

   \section{Photon Doppler Velocimetry, spectrograms and radiative transfer}\label{pdv}

\begin{figure}[!htb]
     \centering
     \includegraphics[width=0.80\linewidth]{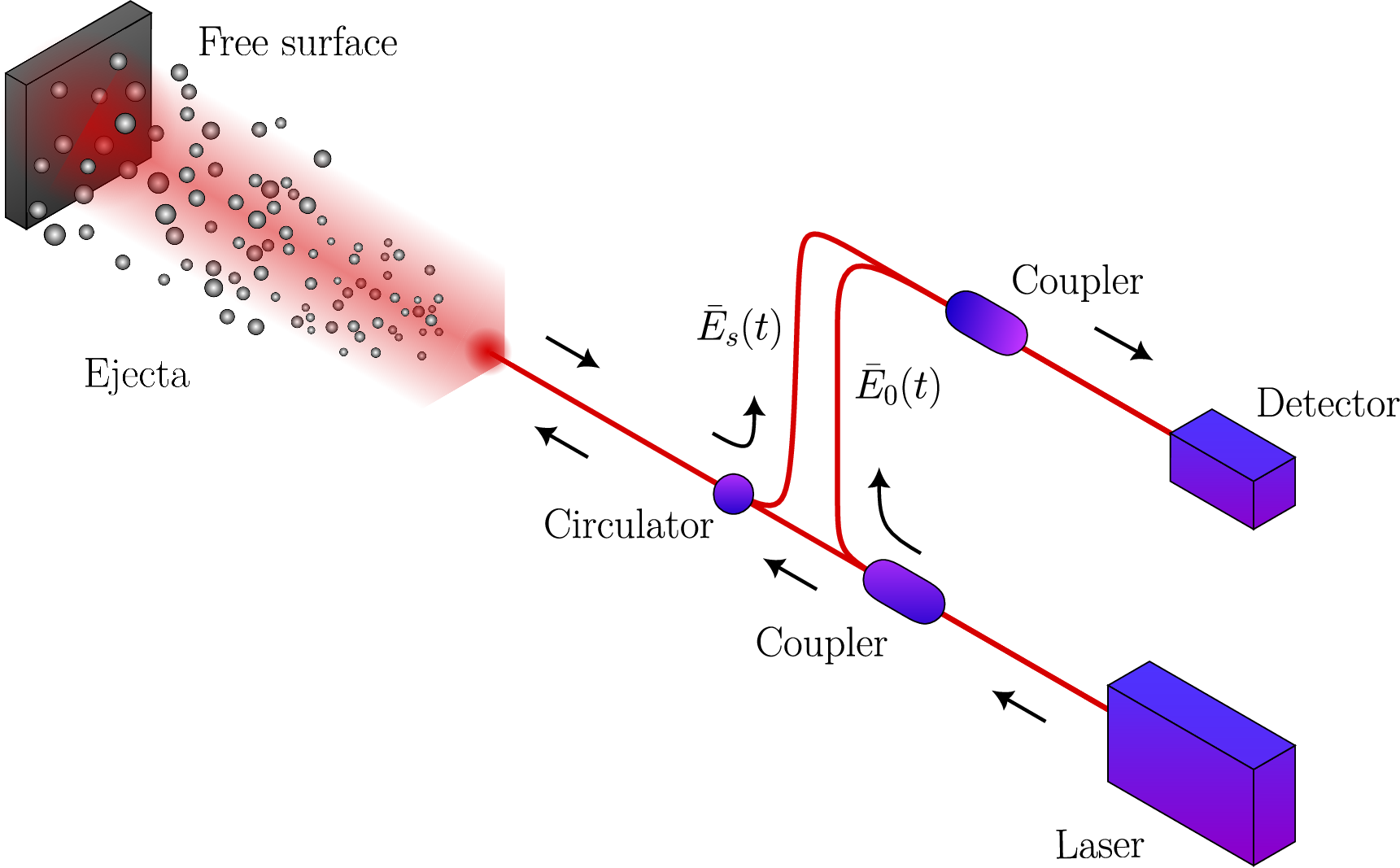}
     \caption{Schematic representation of a typical shock-loaded experiment with a PDV setup.
     The probe illuminates the ejecta and the free surface with a highly collimated laser beam
     (typical numerical aperture of $\theta_p=\SI{4.2}{\milli rad}$ and pupil size $\phi_p =
     \SI{1.6}{\milli m}$). The backscattered field is collected by the probe acting as the
     measuring arm and interferes with the reference arm at the detector. The beating signal is
     registered with a high bandwidth oscilloscope before being analyzed.}
     \label{setup_pdv}
   \end{figure}

   Photon Doppler Velocimetry is an interferometric
   technique~\cite{strand_compact_2006,mercier_photonic_2006} where a probing laser field at
   frequency $\omega_0$ is sent through an optical fiber towards a cloud of moving particles
   ejected from a free surface. As seen in Fig.~\ref{setup_pdv}, light is then scattered by this
   ejecta and slightly shifted in frequency due to the Doppler effect before being partially
   collected in reflection by the same fiber. This illumination and collection geometry was
   historically designed to probe the free surface velocity, but it is now also used to study
   ejecta. The collected field interferes at the detector with a reference field, at $\omega_0$,
   resulting in a beating signal $\mathcal{I}(t)$ at the detector. This heterodyne detection
   ensures a good signal to noise ratio at optical frequencies.  This signal can be written
   \begin{equation}\label{useful_detected_signal}
      \mathcal{I}(t)=2\Re\left[\bar{\bm{E}}_s(\bm{r},t)\cdot\bar{\bm{E}}_0^*(\bm{r},t)\right],
   \end{equation}
   with $\bar{\bm{E}}_s(\bm{r},t)$ [respectively $\bar{\bm{E}}_0(\bm{r},t)$] the analytic signal
   associated to the scattered (respectively reference) field, $\bm{r}$ being the position of
   the probe and the superscript $*$ denoting the complex conjugate.

   In post-treatment, a short-term Fourier transform is applied defining the spectrogram
   $S(t,\Omega)$ as
   \begin{equation}\label{spectrogram}
      S(t,\Omega)=\left|\int \mathcal{I}(\tau)w(\tau-t)\exp(i\Omega\tau)\ud \tau\right|^2
   \end{equation}
   where $w(t)$ is a gate function of typical width $T_w$ such that $\int w(t)\ud t=T_w$. In
   Eq.~(\ref{spectrogram}), it can be easily shown that $\Omega$ corresponds to the Doppler
   shift experienced by light when propagating inside the ejecta. Time $t$ corresponds to
   different instants in the ejecta's expansion. A spectrogram is therefore the result of a
   time-frequency analysis of the field scattered by the ejecta. Note that in the case of single
   backscattering, $\Omega=(4\pi)v/\lambda=2k_0v$ where $v$ is the velocity of the particles and
   $k_0=2\pi/\lambda$ is the wavevector. In this particular case, a spectrogram can therefore be
   used to trace the particle velocity distribution as a function of time.

   In the multiple scattering regime, however, interpretation is more complex and a more refined
   model is required to account for the spectrograms obtained experimentally. This model is
   derived in detail in Ref.~\onlinecite{don_jayamanne_characterization_2024} and we summarize
   here the main steps since they will serve as building blocks for obtaining the diffusion
   equation derived in Sec.~\ref{diffusion}. We first need a quantity to describe light
   propagation in disordered media. A common one, also derived from a time frequency analysis,
   is the specific intensity. It is defined as the space and time Wigner transform of the field
   and is given
   by~\cite{barabanenkov_spectral_1969,rytov_principles_1989,apresyan_radiation_1996,carminati_principles_2021}
   \begin{multline}\label{specific_intensity}
      \delta(k-k_r)I(\bm{r},\bm{u},t,\omega)
   \\
      =\int
      \bra\bar{E}\left(\bm{r}+\frac{\bm{\uprho}}{2},t+\frac{\tau}{2}\right)
      \bar{E}^*\left(\bm{r}-\frac{\bm{\uprho}}{2},t-\frac{\tau}{2}\right)\ket
   \\
      \exp(-ik\bm{u}\cdot\bm{\uprho}+i\omega \tau)\ud \bm{\uprho}\ud \tau
   \end{multline}
   where $k_r$ is the real part of the effective wavevector in the scattering medium, and
   $\bra\cdots\ket$ denotes a statistical average over an ensemble of realizations of the
   disorder (\ie, the scatterers' position). The specific intensity $I(\bm{r},\bm{u}, t,
   \omega)$ is defined from the field-field correlation function. In a radiometric picture, it
   can also be seen as a radiative flux at position $\bm{r}$, time $t$ and frequency $\omega$,
   propagating along direction $\bm{u}$. The time $\tau$ plays the role of the correlation time.
   From various considerations on the time scales involved, we can show the spectrogram can be
   written in terms of the specific intensity, and takes the
   form~\cite{don_jayamanne_characterization_2024}
   \begin{equation}\label{spectrogram_specific_intensity_approx}
      S(t,\Omega)\propto\int_{\theta_p}
      \left[I(\bm{r},\bm{u},t,\omega_0+\Omega)+I(\bm{r},\bm{u},t,\omega_0-\Omega)\right]
      \bm{u}\cdot\bm{n}\ud\bm{u}.
   \end{equation}
   where $\theta_p$ is the extremely narrow cone of collection of the probe around $\bm{n}$, the
   direction normal to the probe's pupil. Typically, we have $\theta_p=\SI{4.2}{\milli rad}$.
   $\ud\bm{u}$ means integration over the solid angle. The key point here is that the
   spectrogram at time $t$ and Doppler frequency $\Omega$, which was initially introduced as a
   very specific time-frequency analysis of a beating signal, now appears as a relatively
   intuitive angular integral of the specific intensity at time $t$ and frequencies
   $\omega_0\pm\Omega$. This result makes the specific intensity the quantity of choice to
   describe spectrograms of ejecta.
 
   Next, we need an equation governing the evolution of the specific intensity in ejecta. It is
   given by a generalized RTE~\cite{don_jayamanne_characterization_2024}
   \begin{multline}\label{rte}
      \left[\frac{1}{v_E(\bm{r},t,\omega)}\frac{\partial}{\partial t}+\bm{u}\cdot\bm{\nabla}_{\bm{r}}+\frac{1}{\ell_e(\bm{r},
      t,\omega)}\right]
      I(\bm{r},\bm{u},t,\omega)\\
      =\frac{1}{\ell_s(\bm{r},t,\omega)}\int_{4\pi} p(\bm{r},\bm{u},\bm{u}',t,\omega,\omega')
      I(\bm{r},\bm{u}',t,\omega')\ud\bm{u}'\frac{\ud \omega'}{2\pi},
   \end{multline}
   with $v_E$ the energy velocity, $\ell_e$ the extinction mean free path, $\ell_s$ the
   scattering mean free path and $p$ the generalized phase function. In the phase function,
   $\bm{u}'$ and $\omega'$ are the incident direction and frequency, respectively, and $\bm{u}$
   and $\omega$ are the scattered direction and frequency, respectively. Equation~(\ref{rte})
   takes into account the specificities of ejecta such as statistical inhomogeneity,
   polydispersity and the motion of the particles in the ejecta.
   
   We assume in the following that the ejecta is composed of spherical particles of different
   radii with an inhomogeneous number density depending on position and time. In this case, the
   extinction mean free path $\ell_e$ is given by
   \begin{equation}\label{l_e}
      \frac{1}{\ell_e(\bm{r},t,\omega)}
         =\int\rho(\bm{r},t) \sigma_{e}(a,\omega)h(\bm{r},t,a)\ud a,
   \end{equation}
   where $\rho(\bm{r},t)$ is the number density of particles at position $\bm{r}$ and time $t$
   and $\sigma_{e}(a,\omega)$ is the extinction cross-section of a particle with radius $a$ at
   frequency $\omega$. $h(\bm{r},t,a)$ is the size distribution at position $\bm{r}$ and time
   $t$. The scattering mean free path $\ell_s$ and the generalized phase function $p$ are
   defined as
   \begin{multline}\label{l_s_p}
      \frac{1}{\ell_s(\bm{r},t,\omega)} p(\bm{r},\bm{u},\bm{u}',t,\omega,\omega')
      =\int\rho(\bm{r},t)\sigma_{s}(a,\omega)p(a,\bm{u}\cdot\bm{u}^\prime,\omega)
   \\\times
      2\pi\delta\left[\omega'-\omega-k_r(\bm{u}'-\bm{u})\cdot\bm{v}\right]
      f(\bm{r},t,a,\bm{v})\ud a\ud\bm{v},
   \end{multline}
   where $\sigma_{s}(a,\omega)$ and $p(a,\bm{u}\cdot\bm{u}^\prime,\omega)$ are the scattering
   cross-section and the phase function of a particle with radius $a$ at frequency $\omega$
   respectively. The size-velocity distribution at position $\bm{r}$ and time $t$ is given by
   $f(\bm{r},t,a,\bm{v})$ with $h(\bm{r},t,a)=\int f(\bm{r},t,a,\bm{v}) \ud\bm{v}$. We note that
   the phase function $p(a,\bm{u}\cdot\bm{u}^\prime,\omega)$ depends only on the dot product
   $\bm{u}\cdot\bm{u}^\prime$ for a spherical particle. It is given by
   \begin{equation}
      p(a,\bm{u}\cdot\bm{u}^\prime,\omega)=\frac{1}{\sigma_{s}(a,\omega)}
      \frac{\ud\sigma_{s}(a,\bm{u}\cdot\bm{u}',\omega)}{\ud\bm{u}},
   \end{equation}
   where the differential scattering cross section
   $\ud\sigma_{s}(a,\bm{u}\cdot\bm{u}',\omega)/\ud\bm{u}$ corresponds to the radiation pattern.
   It is proportional to the part of radiated power coming from incoming direction
   $\bm{u}^\prime$ and scattered along out going direction $\bm{u}$. Its integration over all
   possible directions gives the total scattering cross section. With this definition, the phase
   function is normalized as
   \begin{equation}
      \int_{4\pi} p(\bm{r},\bm{u},\bm{u}',t,\omega,\omega')\ud \bm{u}'\frac{\ud\omega'}{2\pi}=1.
   \end{equation}
   Thus, integrating Eq.~(\ref{l_s_p}) over $\bm{u}$, the scattering mean free-path reads
   \begin{equation}\label{l_s}
      \frac{1}{\ell_s(\bm{r},t,\omega)}
         =\int \rho(\bm{r},t) \sigma_{s}(a,\omega)h(\bm{r},t,a)\ud a.
   \end{equation}
   We also introduce the absorption mean free path $\ell_a(\bm{r},t)$ such that
   \begin{equation}
      \frac{1}{\ell_a(\bm{r},t,\omega)}=\frac{1}{\ell_e(\bm{r},t,\omega)}-\frac{1}{\ell_s(\bm{r},t,\omega)}.
   \end{equation}
   Finally, since we assume nonresonant scattering, the energy velocity $v_E$ is given by
   $v_E(\bm{r},t,\omega)=c/n_r(\bm{r},t,\omega)$ where $n_r$ is the real part of the effective
   optical index of the disordered medium (\ie, $k_r=n_rk_0$). The detailed physical
   interpretation of the generalized RTE as well as the derivation of the extinction mean free
   path, the scattering mean free path and phase function expressions are discussed in greater
   details in Ref.~\onlinecite{don_jayamanne_characterization_2024}. In practice, the transport
   equation~(\ref{rte}) can be solved numerically using a Monte Carlo scheme allowing to compute
   PDV spectrograms~\cite{don_jayamanne_characterization_2024}.
 
   In a standard experiment, we can assume that the optical properties of an ejecta are
   statistically invariant along the transverse directions and it is common to define the
   optical thickness $b_s$ in an inhomogeneous ejecta as
   \begin{equation}
      b_s=\int_0^L\frac{\ud z}{\ell_s(z)}
   \end{equation}
   where $z$ denotes the position along the ejection direction and $L$ the size of the ejecta
   along this direction. For the example treated in
   Ref.~\onlinecite{don_jayamanne_characterization_2024} we found $b_s=42$, which means that
   light interacts with the ejecta in the deep multiple scattering regime. For such optical
   thicknesses, it seems reasonable to investigate the possibility to use the diffusion
   approximation to describe the spectrograms, instead of the full RTE. This is the subject of
   the following sections.
 
   \section{Derivation of a diffusion equation for light in ejecta}\label{diffusion}
 
   There are different ways to derive a diffusion equation starting from the RTE. One can
   introduce modes of the transport equation and define the diffusion mode as the mode surviving
   at the largest times and depths, a method also known as the singular eigenfunctions
   approach~\cite{elaloufi_definition_2003,pierrat_photon_2006}. One can also perform a direct
   asymptotic analysis of the RTE at large
   scales~\cite{larsen_asymptotic_1974,carminati_principles_2021}. Finally, an angular expansion
   over spherical harmonics of the specific intensity can be also be introduced. A diffusion
   equation is then obtained by keeping only the first two terms, a method known as the P1
   approximation~\cite{furutsu_diffusion_1994,durduran_does_1997,ishimaru_wave_1997}. Here we
   will make use of the angular expansion approach (in the steps of the derivation in
   Ref.~\onlinecite{ishimaru_wave_1997}) since it is the easiest to use in the case of
   statistical inhomogeneous and dynamic media such as ejecta. This idea of introducing a
   diffusion equation for dynamic media is not new. Several models have already been derived
   with different
   techniques~\cite{pine_diffusing-wave_1990,berne_dynamic_2000,scheffold_diffusing-wave_2001,pierrat_influence_2008,pierrat_transport_2008}
   and have been used extensively to study light transport in
   colloids~\cite{harden_recent_2001}, cold atomic gases~\cite{labeyrie_radiation_2005} or
   biological tissues~\cite{klose_light_2006}. We will draw inspiration out of all these works
   in the following. In particular, a fairly standard approach is to derive a diffusion equation
   for a static complex medium to determine the path-length probability density $P(s)$ and then
   include the dynamics of the medium afterwards. Here we rigorously derive the diffusion
   equation from the RTE taking into account in the first place the scatterer displacements and
   all other important ejecta characteristics.
 
   \subsection{Transport equation in the time domain}\label{transport}
 
   We start by considering the quasi-homogeneous and inelastic RTE given in Eq.~(\ref{rte}) in
   which we can neglect the time derivative since the illumination in the experiment is
   monochromatic (steady-state) and the transit time for light within the scattering cloud
   remains short compared to the evolution time scale of the cloud (quasi-static approximation).
   Therefore, the time variable $t$ becomes a parameter and we choose to drop it for the sake of
   simplicity. In addition, since the Doppler shifts around the illumination frequency
   $\omega_0$ are small compared to the typical spectral variation scales of extinction and
   scattering cross-sections, the frequency $\omega$ can be fixed at $\omega_0$ in
   $\sigma_e(a,\omega)$, $\sigma_s(a,\omega)$ and in $p(a,\bm{u}\cdot\bm{u}^\prime,\omega)$ and
   we also drop this dependence for the sake of simplicity. Plugging the expression of the phase
   function, and with the simplifications above, the RTE in Eq.~(\ref{rte}) can be rewritten
   \begin{multline}\label{rte_raw}
      \left[\bm{u}\cdot\bm{\nabla}_{\bm{r}}+\frac{1}{\ell_e(\bm{r})}\right]I(\bm{r},\bm{u},\omega)
      =\rho(\bm{r}) \int \sigma_s(a)p(a,\bm{u}\cdot\bm{u}^\prime)
   \\\times
      2\pi\delta\left[\omega^\prime-\omega-k_r(\bm{u}^\prime-\bm{u})\cdot\bm{v}\right]
      f(\bm{r},a,\bm{v})
   \\\times
      I(\bm{r},\bm{u}^\prime,\omega^\prime)
      \ud\bm{u}^\prime\frac{\ud \omega^\prime}{2\pi}\ud a \ud \bm{v}.
   \end{multline}
   This form of the transport equation is not yet fully adapted to the derivation of a diffusion
   equation because the presence of an integral over frequencies needs to be handled correctly.
   To this end, we perform a Fourier transform with respect to $\omega-\omega_0$. This leads to
   \begin{multline}\label{field_decorrelation_rte_raw}
      \left[\bm{u}\cdot\bm{\nabla}_{\bm{r}}+\frac{1}{\ell_e(\bm{r})}\right]I(\bm{r},\bm{u},\tau)
   \\
      =\rho(\bm{r}) \int_{4\pi} \sigma_s(a)p(a,\bm{u}\cdot\bm{u}^\prime)e^{ik_r(\bm{u}^\prime-\bm{u})\cdot\bm{v}\tau}
   \\\times
      f(\bm{r},a,\bm{v}) I(\bm{r},\bm{u}^\prime,\tau) \ud\bm{u}^\prime\ud a \ud \bm{v},
   \end{multline}
   with
   \begin{equation}\label{field_correlation}
      I(\bm{r},\bm{u},\tau)=\int I(\bm{r},\bm{u},\omega)e^{-i(\omega-\omega_0)\tau}\frac{\ud \omega}{2\pi},
   \end{equation}
   $\omega_0$ being a parameter in this expression. The factor
   $e^{ik_r(\bm{u}^\prime-\bm{u})\cdot\bm{v}\tau}$ appearing in
   Eq.~(\ref{field_decorrelation_rte_raw}) shows that inelastic scattering results in the
   appearance of a dephasing in the time domain. As the specific intensity
   $I(\bm{r},\bm{u},\tau)$ propagates through the medium, it accumulates phase shifts which at
   long correlation times eventually leads to full decorrelation.

   The next step consists in defining effective scattering properties to make
   Eq.~(\ref{field_decorrelation_rte_raw}) similar to the standard RTE and then follow as
   closely as possible the steps of the standard derivation of the diffusion approximation
   presented in Ref.~\onlinecite{ishimaru_wave_1997}. The effective scattering mean free path
   and phase function are respectively defined as
   \begin{align}
      \begin{split}
         \label{l_tilde}
         \frac{1}{\tilde{\ell}_s(\bm{r},\bm{u}^\prime, \tau)}
         ={} & \rho(\bm{r})\int_{4\pi} \sigma_s(a) p(a,\bm{u}\cdot\bm{u}^\prime)
      \\
         & \times
            e^{ik_r(\bm{u}^\prime-\bm{u})\cdot\bm{v}\tau}f(\bm{r},a,\bm{v})\ud a \ud \bm{v}\ud\bm{u},
      \end{split}
   \\
       \begin{split}
         \label{p_tilde}
         \tilde{p}(\bm{r},\bm{u},\bm{u^\prime},\tau)
         ={} & \tilde{\ell}_s(\bm{r},\bm{u}^\prime, \tau)\rho(\bm{r})\int \sigma_s(a)
         p(a,\bm{u}\cdot\bm{u}^\prime)
      \\
         & \times
            e^{ik_r(\bm{u}^\prime-\bm{u})\cdot\bm{v}\tau}f(\bm{r},a,\bm{v})\ud a \ud \bm{v}.
      \end{split}
   \end{align}
   We note that in the most general case, $\tilde{\ell}_s$ depends on the incoming direction
   $\bm{u^\prime}$. However, it reduces to the standard scattering mean free path when
   $|\bm{v}|=0$ (static scatterers) or when the statistical distribution of velocity is
   isotropic. With these definitions, Eq.~(\ref{field_decorrelation_rte_raw}) becomes
   \begin{multline}\label{field_decorrelation_rte}
      \left[\bm{u}\cdot\bm{\nabla}_{\bm{r}}+\frac{1}{\ell_e(\bm{r})}\right]
      I(\bm{r},\bm{u},\tau)
      =\int_{4\pi}\frac{1}{\tilde{\ell}_s(\bm{r},\bm{u}^\prime,\tau)} \tilde{p}(\bm{r},\bm{u},\bm{u^\prime},\tau)
   \\\times
      I(\bm{r},\bm{u}^\prime,\tau)
      \ud\bm{u}^\prime.
   \end{multline}
 
\begin{figure}[!htb]
      \centering
      \includegraphics[width=0.8\linewidth]{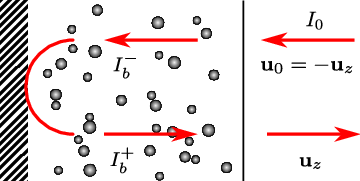}
      \caption{Illustration of the two ballistic components existing in an ejecta illuminated by
      a plane-wave in the direction $\bm{u}_0=-\bm{u}_z$. The free surface is on the left.}
      \label{ballistic}
   \end{figure}

   We now split the specific intensity into its ballistic $I_b$ and diffuse $I_d$ parts. Only
   the diffuse component will be driven by a diffusion equation, with the ballistic component
   appearing in a source term. As pictured in Fig.~\ref{ballistic} and explained in greater
   details in Sec.~\ref{geometry}, in an ejecta geometry, the medium is illuminated along
   direction $-\bm{u}_z$ and the ballistic intensity gets backscattered by the free surface
   along $\bm{u}_z$. To account for both contributions, the specific intensity in the temporal
   domain reads
   \begin{equation}\label{specific_intensity_decomposition}
      I(\bm{r},\bm{u},\tau)=I_d(\bm{r},\bm{u},\tau)+\delta(\bm{u}+\bm{u}_z)I_b^-(\bm{r},\tau)+\delta(\bm{u}-\bm{u}_z)I_b^+(\bm{r},\tau)
   \end{equation}
   where $I_b^-$ is the incident ballistic contribution propagation towards negative $z$ and
   $I_b^+$ is the reflected ballistic contribution propagation towards positive $z$.
   
   Inserting this decomposition into Eq.~(\ref{field_decorrelation_rte}) yields a form of the
   latter with distributions and functions on either side of the equality. The distribution
   terms are then equalized to obtain a  transport equation for $I_b^-(\bm{r},\tau)$ and
   $I_b^+(\bm{r},\tau)$ given by
   \begin{equation}\label{decorrelation_ballistic}
      \left[\pm\bm{u}_z\cdot\bm{\nabla}_{\bm{r}}+\frac{1}{\ell_e(\bm{r})}\right]I_b^\pm(\bm{r},\tau)=0.
   \end{equation}
   In a similar way, function terms are equalized to obtain a transport equation for the diffuse
   intensity $I_d(\bm{r},\bm{u},\tau)$ which reads
   \begin{multline}\label{decorrelation_diffuse}
      \left[\bm{u}\cdot\bm{\nabla}_{\bm{r}}+\frac{1}{\ell_e(\bm{r})}\right]I_d(\bm{r},\bm{u},\tau)
   \\
      =\frac{1}{\tilde{\ell}_s(\bm{r},-\bm{u}_z,\tau)} \tilde{p}(\bm{r},\bm{u},-\bm{u}_z,\tau)I_b^-(\bm{r},\tau)
   \\   
      +\frac{1}{\tilde{\ell}_s(\bm{r},\bm{u}_z,\tau)} \tilde{p}(\bm{r},\bm{u},\bm{u}_z,\tau)I_b^+(\bm{r},\tau)
   \\
      +\int_{4\pi} \frac{1}{\tilde{\ell}_s(\bm{r},\bm{u}^\prime,\tau)}\tilde{p}(\bm{r},\bm{u},\bm{u^\prime},\tau)
      I_d(\bm{r},\bm{u}^\prime,\tau) \ud\bm{u}^\prime.
   \end{multline}
   Since the ballistic terms $I_b^-$ and $I_b^+$ act as a source terms for $I_d$, solving
   Eq.~(\ref{decorrelation_ballistic}) will be a prerequisite to solve the final diffusion
   equation.

   \subsection{Moments of the transport equation and P1 approximation}
 
   A diffusion equation is usually obtained by means of an energy balance and Fick's or
   Fourier's law. The first is obtained from the zero-order angular moment of the RTE. The
   second is derived from the first-order angular moment. This subsection is dedicated to
   obtaining these two equations under the P1 approximation. The zero-order moment of the RTE is
   obtained by integrating Eq.~(\ref{decorrelation_diffuse}) over the direction $\bm{u}$. This
   leads to
   \begin{multline}\label{zero_order_moment_before_p1}
      \bm{\nabla}_{\bm{r}}\cdot\bm{q}_d(\bm{r},\tau)
      +\frac{v_E(\bm{r})}{\ell_e(\bm{r})}u_d(\bm{r},\tau)
      =\int\frac{1}{\tilde{\ell}_s(\bm{r},\bm{u}^\prime,\tau)}I_d(\bm{r},\bm{u}^\prime,\tau)\ud\bm{u}^\prime
   \\
      +S_0 (\bm{r},\tau),
   \end{multline}
   with the energy current defined as
   \begin{equation}\label{q_d}
      \bm{q}_d(\bm{r},\tau)=\int_{4\pi} I_d(\bm{r},\bm{u},\tau)\bm{u}\ud \bm{u},
   \end{equation}
   the diffuse energy density defined as
   \begin{equation}\label{u_d}
      u_d(\bm{r},\tau)=\frac{1}{v_E(\bm{r})}\int_{4\pi} I_d(\bm{r},\bm{u},\tau)\ud \bm{u},
   \end{equation}
   and the source term defined as
   \begin{equation}
      S_0(\bm{r},\tau)=\frac{1}{\tilde{\ell}_s(\bm{r},-\bm{u}_z,\tau)}I_b^-(\bm{r},\tau)+\frac{1}{\tilde{\ell}_s(\bm{r},\bm{u}_z,\tau)}I_b^+(\bm{r},\tau).
   \end{equation}
   For static disorder (\ie, statistical realizations of the medium where the particles are
   assumed to have a null velocity), Eq.~(\ref{zero_order_moment_before_p1}) reduces to a local
   energy balance law involving only $u_d$ and $\bm{q}_d$. The situation is different here due
   to the angular dependence of the effective scattering mean free path. To get closer to an
   energy balance, which is a prerequisite for obtaining a diffusion equation, it is necessary
   to introduce the P1 approximation. It consists in considering that at large depths and long
   times (the meaning of large will be made quantitative later), the specific intensity is
   almost isotropic. Note that this is achievable only if absorption is weak compared to
   scattering. For ejecta, we typically have $\ell_a/\ell_s\approx 10$ [see
   Ref.~\onlinecite{don_jayamanne_multiple_2024}]. The specific intensity can be written as a
   first-order expansion in spherical harmonics, and takes the form that reads
   \begin{equation}\label{p1_correlation}
      I_d(\bm{r},\bm{u},\tau)
      =\frac{v_E(\bm{r})}{4\pi}u_d(\bm{r},\tau)
      +\frac{3}{4\pi}\bm{q}_d(\bm{r},\tau)\cdot \bm{u}.
   \end{equation}
   We note that the prefactors of each order are fully determined by
   Eqs.~(\ref{q_d})~and~(\ref{u_d})~\cite{furutsu_diffusion_1994,durduran_does_1997,ishimaru_wave_1997}.
 
   Plugging this approximation in Eq.~(\ref{zero_order_moment_before_p1}), the zero-order moment
   of the RTE then reads
   \begin{multline}\label{zero_order_moment_after_p1}
      \bm{\nabla}_{\bm{r}}\cdot\bm{q}_d(\bm{r},\tau)
      +\frac{v_E(\bm{r})}{\ell_e(\bm{r})}u_d(\bm{r},\tau)
      =v_E(\bm{r})M_0(\bm{r},\tau)u_d(\bm{r},\tau)
   \\
      +3\bm{M}_1(\bm{r},\tau)\cdot\bm{q}_d(\bm{r},\tau)
      +S_0(\bm{r},\tau),
   \end{multline}
   with 
   \begin{align}
      M_0(\bm{r},\tau)=\frac{1}{4\pi} \int \frac{1}{\tilde{\ell}_s(\bm{r},\bm{u}^\prime,\tau)}\ud \bm{u}^\prime,
   \\
      \bm{M}_1(\bm{r},\tau)=\frac{1}{4\pi} \int \frac{1}{\tilde{\ell}_s(\bm{r},\bm{u}^\prime,\tau)}\bm{u}^\prime\ud \bm{u}^\prime,
   \end{align}
   the first two angular moments of $1/\tilde{\ell}_s$. While in the regular derivation of the
   diffusion equation $M_0=1/\ell_s$, the presence of $\bm{M}_1(\bm{r},\tau)$ is specific to
   dynamic media with anisotropic velocity distribution.

   Likewise, to obtain the first-order angular moment of the transport equation, we multiply
   Eq.~(\ref{decorrelation_diffuse}) by $\bm{u}$ and integrate over $\bm{u}$. We obtain
   \begin{multline}\label{first_order_moment_before_anisotropy}
      \int\bm{u}\cdot\bm{\nabla}_{\bm{r}}I_d(\bm{r},\bm{u},\tau)\bm{u}\ud\bm{u}+
      \frac{1}{\ell_e(\bm{r})}\bm{q}_d(\bm{r},\tau)
   \\
      =\int \frac{1}{\tilde{\ell}_s(\bm{r},\bm{u}^\prime,\tau)}\tilde{p}(\bm{r},\bm{u},\bm{u^\prime},\tau)I_d(\bm{r},\bm{u}^\prime,\tau) \bm{u}\ud\bm{u}^\prime\ud\bm{u} +\bm{S}_1(\bm{r},\tau),
   \end{multline}
   with the source term
   \begin{multline}
      \bm{S}_1(\bm{r},\tau)=\frac{1}{\tilde{\ell}_s(\bm{r},-\bm{u}_z,\tau)}I_b^-(\bm{r},\tau)\int \tilde{p}(\bm{r},\bm{u},-\bm{u}_z,\tau)\bm{u}\ud\bm{u}
   \\
      +\frac{1}{\tilde{\ell}_s(\bm{r},\bm{u}_z,\tau)}I_b^+(\bm{r},\tau)\int \tilde{p}(\bm{r},\bm{u},\bm{u}_z,\tau)\bm{u}\ud\bm{u}.
   \end{multline}
   As in the usual derivation~\cite{ishimaru_wave_1997}, we define the (effective) anisotropy
   factor as
   \begin{equation}
      \tilde{g}(\bm{r},\bm{u}^\prime,\tau)=\int\tilde{p}(\bm{r},\bm{u},\bm{u}^\prime,\tau)(\bm{u}\cdot\bm{u}^\prime)\ud\bm{u},
   \end{equation}
   which allows us to rewrite the first term in the right-hand side of
   Eq.~(\ref{first_order_moment_before_anisotropy}) as
   \begin{multline}
      \int \frac{1}{\tilde{\ell}_s(\bm{r},\bm{u}^\prime,\tau)}\tilde{p}(\bm{r},\bm{u},\bm{u}^\prime,\tau)I_d(\bm{r},\bm{u}^\prime,\tau) \bm{u}\ud
      \bm{u}^\prime\ud\bm{u}
   \\
      =\int \frac{\tilde{g}(\bm{r},\bm{u}^\prime,\tau)}{\tilde{\ell}_s(\bm{r},\bm{u}^\prime,\tau)}I_d(\bm{r},\bm{u}^\prime,\tau) \bm{u}^\prime\ud\bm{u}^\prime.
   \end{multline}
   Plugging again the P1 approximation given by Eq.~(\ref{p1_correlation}) in this relation, we
   obtain
   \begin{multline}
      \int
      \frac{\tilde{g}(\bm{r},\bm{u}^\prime,\tau)}{\tilde{\ell}_s(\bm{r},\bm{u}^\prime,\tau)}I_d(\bm{r},\bm{u}^\prime,\tau)
         \bm{u}^\prime\ud\bm{u}^\prime
   \\
      =v_E(\bm{r})u_d(\bm{r},\tau)\bm{G}_1(\bm{r},\tau)+\tens{G}_2(\bm{r},\tau)\bm{q}_d(\bm{r},\tau),
   \end{multline}
   with
   \begin{align}
      \bm{G}_1(\bm{r},\tau) & =\int\frac{\tilde{g}(\bm{r},\bm{u}^\prime,\tau)}{4\pi\tilde{\ell}_s(\bm{r},\bm{u}^\prime,\tau)}\bm{u}^\prime\ud\bm{u}^\prime,
   \\
      \tens{G}_2(\bm{r},\tau) & =3\int \frac{\tilde{g}(\bm{r},\bm{u}^\prime,\tau)}{4\pi\tilde{\ell}_s(\bm{r},\bm{u}^\prime,\tau)}\left(\bm{u}^\prime\otimes\bm{u}^\prime\right)\ud\bm{u}^\prime.
   \end{align}
   Again, $\bm{G}_1$ would not appear in the standard derivation of the diffusion approximation,
   and $\tens{G}_2$ would be simply given by $g/\ell_s \tens{I}$ with $\tens{I}$ the unit
   second-rank tensor.
   
   In the following, we assume that the energy velocity $v_E$ depends weakly on the position
   $\bm{r}$ since it is given by the real part of the effective refractive index $n_r$ which can
   be approximated by the air refractive index (\ie $n_r\approx 1$) in a dilute medium such as
   an ejecta. Therefore, we can assume $\bm{\nabla}_{\bm{r}}v_E(\bm{r})\approx0$ and using again the
   P1 approximation, it is straightforward to show that the first term in the left-hand side of
   Eq.~(\ref{first_order_moment_before_anisotropy}) can be rewritten as
   \begin{equation}
      \int \bm{u}\cdot\bm{\nabla}_{\bm{r}}I_d(\bm{r},\bm{u},\tau)\bm{u}\ud\bm{u}=\frac{v_E(\bm{r})}{3}\bm{\nabla}_{\bm{r}} u_d(\bm{r},\tau).
   \end{equation}
   Finally, the first moment of the RTE, Eq.~(\ref{first_order_moment_before_anisotropy}), leads
   to
   \begin{multline}
      \frac{v_E(\bm{r})}{3}\bm{\nabla}_{\bm{r}} u_d(\bm{r},\tau)+\frac{1}{\ell_e(\bm{r})}\bm{q}_d(\bm{r},\tau)
      =\bm{G}_1(\bm{r},\tau)v_E(\bm{r})u_d(\bm{r},\tau)
   \\
      \tens{G}_2(\bm{r},\tau)\bm{q}_d(\bm{r},\tau)+\bm{S}_1(\bm{r},\tau).
   \end{multline}
 
   This equation can be cast in a form similar to a Fick's law
   \begin{multline}\label{fick}
      \bm{q}_d(\bm{r},\tau)
      =\tens{D}(\bm{r},\tau)\left[\vphantom{\frac{3}{v_E(\bm{r})}}
      -\bm{\nabla}_{\bm{r}} u_d(\bm{r},\tau)
      \right.\\\left.
      +3\bm{G}_1(\bm{r},\tau)u_d(\bm{r},\tau)
      +\frac{3}{v_E(\bm{r})}\bm{S}_1(\bm{r},\tau)\right],
   \end{multline}
   with the diffusion tensor
   \begin{equation}
      \tens{D}(\bm{r},\tau)=\frac{v_E(\bm{r})}{3}\left[\frac{\tens{I}}{\ell_e(\bm{r})}-\tens{G}_2(\bm{r},\tau)\right]^{-1}.
   \end{equation}

   \subsection{Diffusion equation}

   The features of ejecta, such as statistical inhomogeneities, anisotropy and particle motion,
   greatly complicate the equations with the emergence of new quantities such as
   $\bm{M}_1(\bm{r},\tau)$ and $\bm{G}_1(\bm{r},\tau)$ and the dependence on $\tau$ in many of
   them. Usually, the temporal decorrelation of the beam appears in the model as of an effective
   absorption term~\cite{pine_diffusing-wave_1990}. This effective absorption is very likely to
   be the dominant contribution here as well, and in the following we choose to neglect the
   $\tau$ dependence in all parameters except in the effective absorption, \ie, the term
   $1/\ell_e-M_0$ in Eq.~(\ref{zero_order_moment_after_p1}). In practice, it means replacing
   $\tilde{\ell}_s$ and $\tilde{p}$ introduced in Eqs.~(\ref{l_tilde})~and~(\ref{p_tilde}) by,
   respectively, $\ell_s$ and $p$. The validity of this approximation will be checked in
   Sec.~\ref{results} by comparison to a full treatment of the problem with the RTE. Under this
   assumption, the source terms and the angular moments of the effective scattering mean free
   path can be simplified. We find that
   \begin{multline}
         S_0(\bm{r},\tau)
         = \frac{1}{\tilde{\ell}_s(\bm{r},-\bm{u}_z,\tau)}I_b^-(\bm{r},\tau)+\frac{1}{\tilde{\ell}_s(\bm{r},\bm{u}_z,\tau)}I_b^+(\bm{r},\tau)
      \\
         \approx \frac{I_b^-(\bm{r},\tau)+I_b^+(\bm{r},\tau)}{\ell_s(\bm{r})},
   \end{multline}
   \begin{multline}
         \bm{S}_1(\bm{r},\tau)
         = \frac{I_b^-(\bm{r},\tau)}{\tilde{\ell}_s(\bm{r},-\bm{u}_z,\tau)}\int \tilde{p}(\bm{r},\bm{u},-\bm{u}_z,\tau)\bm{u}\ud\bm{u}
      \\
         +\frac{I_b^+(\bm{r},\tau)}{\tilde{\ell}_s(\bm{r},\bm{u}_z,\tau)}\int \tilde{p}(\bm{r},\bm{u},\bm{u}_z,\tau)\bm{u}\ud\bm{u}
      \\
         \approx g(\bm{r})\frac{I_b^+(\bm{r},\tau)-I_b^-(\bm{r},\tau)}{\ell_s(\bm{r})}\bm{u}_z,
   \end{multline}
   \begin{align}
      \bm{M}_1(\bm{r},\tau) & = \int \frac{\bm{u}^\prime}{4\pi\tilde{\ell}_s(\bm{r},\bm{u}^\prime,\tau)}\ud \bm{u}^\prime
      \approx \bm{0},
   \\
      \bm{G}_1(\bm{r},\tau) & = \int\frac{\tilde{g}(\bm{r},\bm{u}^\prime,\tau)}{4\pi\tilde{\ell}_s(\bm{r},\bm{u}^\prime,\tau)}\bm{u}^\prime\ud\bm{u}^\prime
      \approx \bm{0},
   \\
      \tens{G}_2(\bm{r},\tau) & = \int 3\frac{\tilde{g}(\bm{r},\bm{u}^\prime,\tau)}{4\pi\tilde{\ell}_s(\bm{r},\bm{u}^\prime,\tau)}(\bm{u}^\prime\otimes\bm{u}^\prime)\ud\bm{u}^\prime
      \approx\frac{g(\bm{r})}{\ell_s(\bm{r})}\tens{I},
   \end{align}
   where $g(\bm{r})$ is $\tilde{g}(\bm{r},\bm{u}',\tau)$ taken at $\tau=0$ which therefore does
   not depend on $\bm{u}'$ anymore. The diffusion tensor can also be simplified, in the form
   \begin{equation}
      \tens{D}(\bm{r},\tau)=\frac{v_E(\bm{r})}{3}
      \left[\frac{\tens{I}}{\ell_e(\bm{r})}-\frac{g(\bm{r})}{\ell_s(\bm{r})}\tens{I}\right]^{-1}
      \approx D(\bm{r})\tens{I},
   \end{equation}
   where, again making use of the weak absorption approximation, the diffusion constant is given
   by
   \begin{equation}\label{diffusion_equation_correlation_d}
      D(\bm{r})=\frac{v_E(\bm{r})\ell_t(\bm{r})}{3},
   \end{equation}
   with
   \begin{equation}\label{diffusion_equation_correlation_l_t}
      \ell_t(\bm{r})=\frac{\ell_s(\bm{r})}{1-g(\bm{r})},
   \end{equation}
   the transport mean free path. We put forward that in this setting the effective absorption
   length remains unmodified and is given by
   \begin{align}\label{l_a_eff}
      \begin{split}
      \frac{1}{\tilde{\ell}_a(\bm{r},\tau)}
      ={} & \frac{1}{\ell_e(\bm{r})}-M_0(\bm{r},\tau)
   \\
      ={} & \frac{1}{\ell_e(\bm{r})}-\frac{\rho(\bm{r})}{4\pi}
      \int \sigma_s(a,\omega_0)
      p(a,\omega_0,\bm{u}\cdot\bm{u}^\prime)
   \\&\times
      e^{ik_r(\bm{u}^\prime-\bm{u})\cdot\bm{v}\tau}
      f(\bm{r},a,\bm{v})\ud a \ud \bm{v}\ud\bm{u}\ud\bm{u}^\prime.
      \end{split}
   \end{align}
   In this formulation, the decorrelation $M_0$ appears as an extra contribution to the
   intrinsic absorption $1/\ell_a$.
 
   With this expression of the effective absorption, the zero-order moment of the RTE [\ie,
   Eq.~(\ref{zero_order_moment_after_p1})] and the Fick's law [\ie, Eq.~(\ref{fick})] can be
   rewritten in the form
   \begin{align}
      \label{moments_RTE_after_weak_absorption_approximation_0}
      \bm{\nabla}_{\bm{r}}\cdot\bm{q}_d(\bm{r},\tau)
      & = -\frac{v_E(\bm{r})}{\tilde{\ell}_a(\bm{r},\tau)}u_d(\bm{r},\tau)+\frac{I_b^-(\bm{r},\tau)+I_b^+(\bm{r},\tau)}{\ell_s(\bm{r})},
   \\\nonumber
      \bm{q}_d(\bm{r},\tau)
      & = -D(\bm{r})\bm{\nabla}_{\bm{r}} u_d(\bm{r},\tau)
      +\frac{g(\bm{r})}{1-g(\bm{r})}
   \\ & \hphantom{=}\times
      \label{moments_RTE_after_weak_absorption_approximation_1}
      \left[I_b^+(\bm{r},\tau)-I_b^-(\bm{r},\tau)\right]\bm{u}_z.
   \end{align}
 
   Combining
   Eqs.~(\ref{moments_RTE_after_weak_absorption_approximation_0})~and~(\ref{moments_RTE_after_weak_absorption_approximation_1}),
   we readily obtain the following diffusion equation for the diffuse energy density
   \begin{equation}\label{diffusion_equation_correlation}
      -\bm{\nabla}_{\bm{r}}\cdot[D(\bm{r})\bm{\nabla}_{\bm{r}}u_d(\bm{r},\tau)]+\frac{v_E(\bm{r})}{\tilde{\ell}_a(\bm{r},\tau)}u_d(\bm{r},\tau)
      =S(\bm{r},\tau),
   \end{equation}
   with the source term
   \begin{multline}\label{diffusion_equation_correlation_source_term}
      S(\bm{r},\tau)=\frac{I_b^-(\bm{r},\tau)+I_b^+(\bm{r},\tau)}{\ell_s(\bm{r})}
   \\
      -\bm{\nabla}_{\bm{r}}\left\{\frac{g(\bm{r})}{1-g(\bm{r})}\left[I_b^+(\bm{r},\tau)-I_b^-(\bm{r},\tau)\right]\right\}\bm{u}_z.
   \end{multline}
   Equation~(\ref{diffusion_equation_correlation}) is the first important result of this work
   and deserves to be commented. First, the quasi-inhomogeneous aspect of the ejecta is encoded
   in the position dependence of the diffusion constant. Second, the dynamics of the particle
   cloud is taken into account in the $\tau$ dependence of the effective absorption mean free
   path. The standard diffusion equation is recovered in the absence of any $\bm{r}$ and $\tau$
   dependence of $D(\bm{r})$ and $\tilde{\ell}_a(\bm{r},\tau)$.
 
   \subsection{Geometry, boundary conditions and source term}\label{geometry}

   To complete this model, we need to determine the source term and the boundary conditions. To
   this end, we need to define the geometry used to represent the ejecta.
 
\begin{figure}[!htb]
      \centering
      \includegraphics[width=0.8\linewidth]{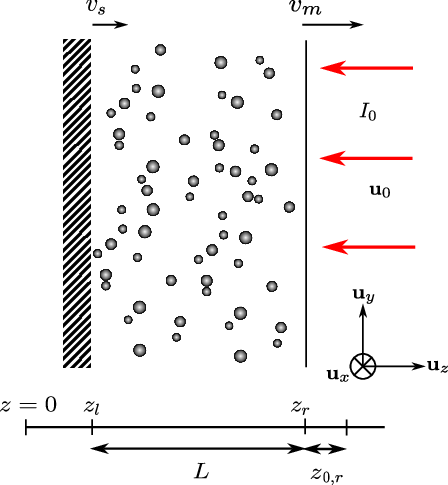}
      \caption{Translation invariant and infinite along $x$-axis and $y$-axis scattering slab of
      length $L$ along its longitudinal $z$-axis. This slab is illuminated from the right by a
      plane wave of intensity $I_0$ and frequency $\omega_0$ propagating in the direction
      $\bm{u}_0 = -\bm{u}_z$. The front of the particle cloud is moving at velocity $v_m$ and
      constitutes the right interface. The boundary condition for the diffuse energy density at
      this interface is given by the usual extrapolation length $z_{0,r}=(2/3)\ell_t(z_r)$.  The
      left interface is the reflective free surface moving at velocity $v_s$. The position $z=0$
      marks the initial position of the free surface and the particle cloud at $t=0$.}
      \label{slab_ejecta}
   \end{figure}

   We consider a translation invariant (infinite) scattering slab along the transverse $x$ and
   $y$ directions, with length $L$ along its longitudinal $z$-axis (see Fig.~\ref{slab_ejecta}).
   This is a well-suited configuration for ejecta generated by a planar shock. The left-hand
   side interface of this slab, located at $z=z_l$, is assumed to be a reflective free surface
   traveling at velocity $v_s\bm{u}_z$. The right-hand side interface is an open interface
   located at $z=z_r$ and traveling at velocity $v_m\bm{u}_z$ (maximum cloud velocity). Since
   position $z=0$ marks the initial position of the free surface and the particle cloud at
   $t=0$, we have
   \begin{equation}
      z_l=v_s t, \quad z_r=v_m t \quad\text{and}\quad L=(v_m-v_s)t.
   \end{equation}
 
   This medium is illuminated by a plane-wave with frequency $\omega_0$ and intensity $I_0$
   propagating along the direction $\bm{u}_0=-\bm{u}_z$. This does not correspond to the real
   illumination of a standard ejecta experiment where an optical fiber is used but it has the
   advantage of preserving translational invariance along $x$ and $y$. Considering perfect
   reflection at the free surface at position $z_l$, the ballistic intensities $I_b^-$ and
   $I_b^+$ governed by Eq.~(\ref{decorrelation_ballistic}) are given by
   \begin{align}
      \label{ballistic_slab_minus}
      I_b^-(z,\tau)&=I_0\exp\left[-\int_{z}^{z_r}\frac{1}{\ell_e(z^\prime)}\ud z^\prime\right],
   \\
      \label{ballistic_slab_plus}
      I_b^+(z,\tau)&=I_b^-(z_l)\exp(-2ik_r v_s\tau)\exp\left[-\int_{z_l}^{z}\frac{1}{\ell_e(z^\prime)}\ud z^\prime\right].
   \end{align}
   We note here that $I_b^-$ does not decorrelate in time, and we can drop the $\tau$
   dependency, while $I_b^+$ decorrelates due to the motion of the free surface. Then the source
   term of the diffusion equation becomes
   \begin{equation}\label{diffusion_equation_correlation_source_term_slab}
      S(z,\tau)=
      \frac{I_b^-(z)+I_b^+(z,\tau)}{\ell_s(z)}
      -\frac{\ud}{\ud z}\left\{\frac{g(z)}{1-g(z)}\left[I_b^+(z,\tau)-I_b^-(z)\right]\right\},
   \end{equation}
   which finally leads to the diffusion equation
   \begin{equation}\label{diffusion_equation_correlation_slab}
      -\frac{\partial}{\partial z}\left[D(z)\frac{\partial}{\partial z}u_d(z,\tau)\right]+\frac{v_E(z)}{\tilde{\ell}_a(z,\tau)}u_d(z,\tau)
      =S(z,\tau).
   \end{equation}

   The final step concerns boundary conditions for the diffuse energy density $u_d$. Determining
   the relevant boundary conditions is not an easy task since $u_d$ does not hold any
   information on direction. The problem can be tackled in different ways. One possibility
   consists in solving the Milne problem~\cite{milne_radiative_1921,case_elementary_1960}. Here
   we choose to stick to the P1 approximation~\cite{ishimaru_wave_1997} which gives similar
   results.
 
   Let us start with the boundary condition at the interface at $z=z_r$. The fact that there is
   no incoming diffuse light means that in terms of specific intensity
   \begin{equation}
      \int_{\bm{u}\cdot\bm{u}_z<0}I_d(z_r,\bm{u},\tau)\bm{u}\cdot\bm{u}_z\ud\bm{u}=0.
   \end{equation}
   Plugging in this expression the P1 approximation and noting that
   Eq.~(\ref{moments_RTE_after_weak_absorption_approximation_1}) becomes
   \begin{equation}\label{fick_slab}
      \bm{q}_d(z,\tau)\cdot \bm{u}_z =-D(z)\frac{\partial}{\partial z}u_d(z,\tau) +\frac{g(z)}{1-g(z)}\left[I_b^+(z,\tau)-I_b^-(z)\right]
   \end{equation}
   in the slab geometry, we find that
   \begin{equation}\label{boundary_right}
      u_d(z_r,\tau)+z_{0,r}\frac{\partial}{\partial z}u_d(z=z_r,\tau)
      =\frac{2g(z_r)\left[I_b^+\left(z_r,\tau\right)-I_b^-\left(z_r\right)\right]}{v_E(z_r)[1-g(z_r)]},
   \end{equation}
   with $z_{0,r}=(2/3)\ell_t(z_r)$. We recover here the usual boundary condition for an open
   interface~\cite{ishimaru_wave_1997,carminati_principles_2021} where $z_{0,r}$ is known as the
   extrapolation length.

   Considering perfect reflection on the free surface at $z=z_l$, we find the relation
   \begin{equation}
      I_d(z_l,\bm{u},\tau)=I_d\left[z_l,\bm{u}-2(\bm{u}\cdot\bm{u}_z)\bm{u}_z,\tau\right]
         \exp[-2ik_{r}v_s(\bm{u}\cdot\bm{u}_z)\tau]
   \end{equation}
   on the diffuse specific intensity for $\bm{u}\cdot\bm{u}_z>0$. The first angular moment of
   the relation leads to an equality of the radiative flux given by
   \begin{multline}\label{flux_fs}
      \int_{\bm{u}\cdot\bm{u}_z>0}I_d(z_l,\bm{u},\tau)\bm{u}\cdot\bm{u}_z\ud\bm{u}
      =-\int_{\bm{u}\cdot\bm{u}_z>0}I_d\left[z_l,\bm{u}-2(\bm{u}\cdot\bm{u}_z)\bm{u}_z,\tau\right]
   \\\times
      \exp[-2ik_{r}v_s(\bm{u}\cdot\bm{u}_z)\tau]
      \left[\bm{u}-2(\bm{u}\cdot\bm{u}_z)\bm{u}_z\right]\cdot\bm{u}_z\ud\bm{u}.
   \end{multline}
   Again, plugging the P1 approximation and making use of Eq.~(\ref{fick_slab}) leads to
   \begin{multline}\label{fs}
      -\left[\frac{1}{2}+\frac{3}{2}b(\tau)\right]\frac{\partial}{\partial z} u_d(z=z_l,\tau)
   \\
      =-\left[\frac{1}{2}+\frac{3}{2}b(\tau)\right]\frac{g(z_l)}{D(z_l)[1-g(z_l)]}\left[I_b^+(z_l)-I_b^-(z_l)\right]
   \\
      +\frac{v_E(z_l)}{D(z_l)}\left[\frac{1}{4}-\frac{a(\tau)}{2} \right] u_d(z_l,\tau)
   \end{multline}
   where
   \begin{align}
       \begin{split}
         a(\tau) ={} & i\frac{1}{2k_rv_s\tau}\exp(-2ik_r v_s\tau)
      \\
         & -i\frac{1}{2k_rv_s\tau}\sinc(k_r v_s \tau)\exp(-ik_r v_s\tau),
      \end{split}
   \\
      b(\tau) ={} & i\frac{1}{2k_rv_s\tau}\exp(-2ik_r v_s\tau)-i\frac{a(\tau)}{k_r v_s \tau}.
   \end{align}
   We note that when $\tau=0$ or equivalently $v_s=0$, we simply have $a=1/2$ and $b=1/3$. From
   Eq.~(\ref{fick_slab}), we find that $\bm{q}_d(z_l,\tau=0)\cdot\bm{u}_z=0$ which is consistent
   with the expected result for perfect reflection (no radiative flux).

   \subsection{Collected flux in reflection}

   In order to describe real PDV spectrograms using the diffusion equation, we first need to
   express the flux collected by a PDV probe as a function of the ballistic intensity $I_b^+$
   and of the diffuse energy density.
 
   In terms of the specific intensity, the flux collected by the probe is
   \begin{equation}
      \Phi_p(\tau)=S_p\left[ I_b^+(z_r,\tau)+\int_{\theta_p} I_d(z_r,\bm{u},\tau)(\bm{u}\cdot\bm{u}_z)\ud\bm{u}\right],
   \end{equation}
   where $S_p$ is the surface of the probe and $\theta_p$ is the angle of collection. Using the
   P1 approximation given by Eq.~(\ref{p1_correlation}) and the definition of the numerical
   aperture $\mathrm{NA}=n_r\sin{\theta_p}$, with $n_r$ the real part of the effective
   refractive index, the flux collected by the probe can be rewritten as
   \begin{multline}
      \Phi_p(\tau)=S_p I_b^+(z_r,\tau)+S_p\frac{v_E}{4}\frac{\mathrm{NA}^2}{n_r^2} u_d(z_r,\tau)
   \\
      +S_p\frac{1}{2}\left[1-\left(1-\frac{\mathrm{NA}^2}{n_r^2}\right)^{3/2}\right][\bm{q}_d(z_r,\tau)\cdot\bm{u}_z].
   \end{multline}
   Using Eqs.~(\ref{fick_slab})~and~(\ref{boundary_right}) to write the flux as a function of
   the diffuse energy density, we obtain
   \begin{multline}\label{phi_p}
      \Phi_p (\tau)=S_p I_b^+(z_r,\tau)
   \\
      +S_p \frac{v_E}{4}\left\{\frac{\mathrm{NA}^2}{n_r^2} +\left[1-\left(1-\frac{\mathrm{NA}^2}{n_r^2}\right)^{3/2}\right]\right\}u_d(z_r,\tau).
   \end{multline}
   Since the specific intensity is assumed to be almost isotropic, we clearly see that the
   $\theta_p$ dependence of the collected flux factorizes. In the following, the expressions of
   $\Phi_p(\tau)$ and of the spectrogram $S(\Omega)$ will both be useful. From
   Eqs.~(\ref{spectrogram_specific_intensity_approx})~and~(\ref{field_correlation}), we see that
   the expression of the spectrogram takes the form
   \begin{equation}\label{spectrogram_phi_p}
      S(\Omega)\propto \int \Phi_p(\tau)\cos(\Omega\tau)\ud \tau.
   \end{equation}

   In this section, we have derived a diffusion equation relevant for the description of light
   propagation in ejecta, accounting for the statistical inhomogeneities and the motion of the
   scatterers. We have completed this model by adding appropriate boundary conditions and
   defining the collected flux. The next step is now to test the validity of the diffusion model
   against full RTE simulations, which is the objective of the next section.
 
   \section{Potential and limitation of the diffusion model}\label{results}

   This section is dedicated to comparing the results obtained by a numerical resolution of the
   RTE and the solution of the diffusive model. The main idea is to start with a simple
   situation in which the physical parameters are chosen such that the solution of the diffusion
   approximation is expected to match the full numerical solution of the RTE. Then, one by one,
   we will increase the complexity by adding different ejecta characteristics to reach a
   situation corresponding to a real ejecta. This process will allow us to check step by step
   the validity of the diffusion model.

   In the following, the numerical resolution of the RTE is performed using a Monte Carlo
   scheme, as presented in Ref.~\onlinecite{don_jayamanne_characterization_2024}. Note that this
   scheme can be adapted to either compute the specific intensity in the $\tau$-domain [\ie,
   $I(\bm{r},\bm{u},\tau)$] or in the $\omega$-domain [\ie, $I(\bm{r},\bm{u},\omega)$] directly
   without requiring  a Fourier transform which is more computationally efficient. The diffusion
   model given by Eq.~(\ref{diffusion_equation_correlation_slab}) together with the boundary
   conditions given by Eqs.(\ref{boundary_right}) and~(\ref{fs}) is solved using a
   ﬁnite-difference scheme. Once the diffuse energy density $u_d(z,\tau)$ has been computed, the
   ﬂux collected by the probe, the spectrogram and other derived quantities can be obtained
   directly. As an order of magnitude, the computation of a spectrogram using the Monte Carlo
   scheme requires about \num{1e8} photons and takes about \SI{8}{h} on a computing cluster
   using 24 48-core Intel Xeon Gold 5220R, each clocked at \SI{2.2}{GHz}, while the solution of
   the diffusion equation takes only \SI{1}{h} on a regular laptop using a 4-core Intel Core
   i7-8665U CPU clocked at \SI{1.90}{GHz}.  This clearly illustrates the interest of using a
   diffusion model whenever it is relevant.

   \subsection{Statistically homogeneous medium with isotropic velocity
   distribution}\label{pine}
 
   We first consider the simple case of a statistically homogeneous particle cloud, the number
   density of which is given by
   \begin{equation}
      \rho=\frac{M_s}{m_p L},
   \end{equation}
   where $M_s$ is the total ejected surface mass and $m_p=4/3\pi \rho_\text{Sn}a_0^3$ is the
   mass of a single particle, $\rho_\text{Sn}$ being the volume density of tin. We recall that
   $L$ is the cloud thickness (see Fig.~\ref{slab_ejecta}). We also choose an isotropic Gaussian
   velocity distribution
   \begin{equation}\label{pine_velocity}
      j(\bm{v})=\frac{1}{\sigma_v^2(2\pi)^{3/2}}\exp\left(-\frac{|\bm{v}|^2}{2\sigma_v^2}\right).
   \end{equation}
   We finally consider a mono disperse medium such that the particle size distribution is given
   by
   \begin{equation}
      h(a)=\delta(a-a_0),
   \end{equation}
   where $a_0$ is the radius of the particles. In this setting, the size-velocity distribution
   is simply
   \begin{equation}
      f(z,a,\bm{v})=h(a)j\left(\bm{v}\right).
   \end{equation}
   Additionally, we neglect the role of the free surface and assume open boundary conditions at
   both $z_l$ and $z_r$ taking the form
   \begin{equation}\label{boundary_left_pine}
      u_d(z_l,\tau)-z_{0,l}\frac{\partial}{\partial z}u_d(z=z_l,\tau)
      =\frac{2gI_b^-\left(z_l\right)}{v_E(1-g)},
   \end{equation}
   \begin{equation}\label{boundary_right_pine}
      u_d(z_r,\tau)+z_{0,r}\frac{\partial}{\partial z}u_d(z=z_r,\tau)
      =-\frac{2gI_b^-\left(z_r\right)}{v_E(1-g)},
   \end{equation}
   with $z_{0,l}=z_{0,r}=(2/3)\ell_t$. We note that $I_b^+=0$ in the absence of the free
   surface.

   Making $\tilde{\ell}_a$ given in Eq.~(\ref{l_a_eff}) explicit using the velocity distribution
   given in Eq.~(\ref{pine_velocity}) leads to
   \begin{equation}\label{l_a_eff_pine}
      \frac{1}{\tilde{\ell}_a(\tau)}
      =\frac{1}{\ell_e}-\frac{2\pi}{\ell_s}\int \exp\left[-\sigma_v^2k_r^2\tau^2(1-\mu)\right]p(a_0,\mu)\ud\mu,
   \end{equation}
   where $\mu=\bm{u}\cdot\bm{u}^\prime$. We see that at long correlation times, $\lim_{\tau
   \to+\infty} 1/\tilde{\ell}_a(\tau)=1/\ell_e$. The integral over $\mu$ is computed numerically
   and this configuration will be used as a reference in the following.
 
\begin{figure*}[!ht]
      \centering
      \includegraphics[width=0.80\linewidth]{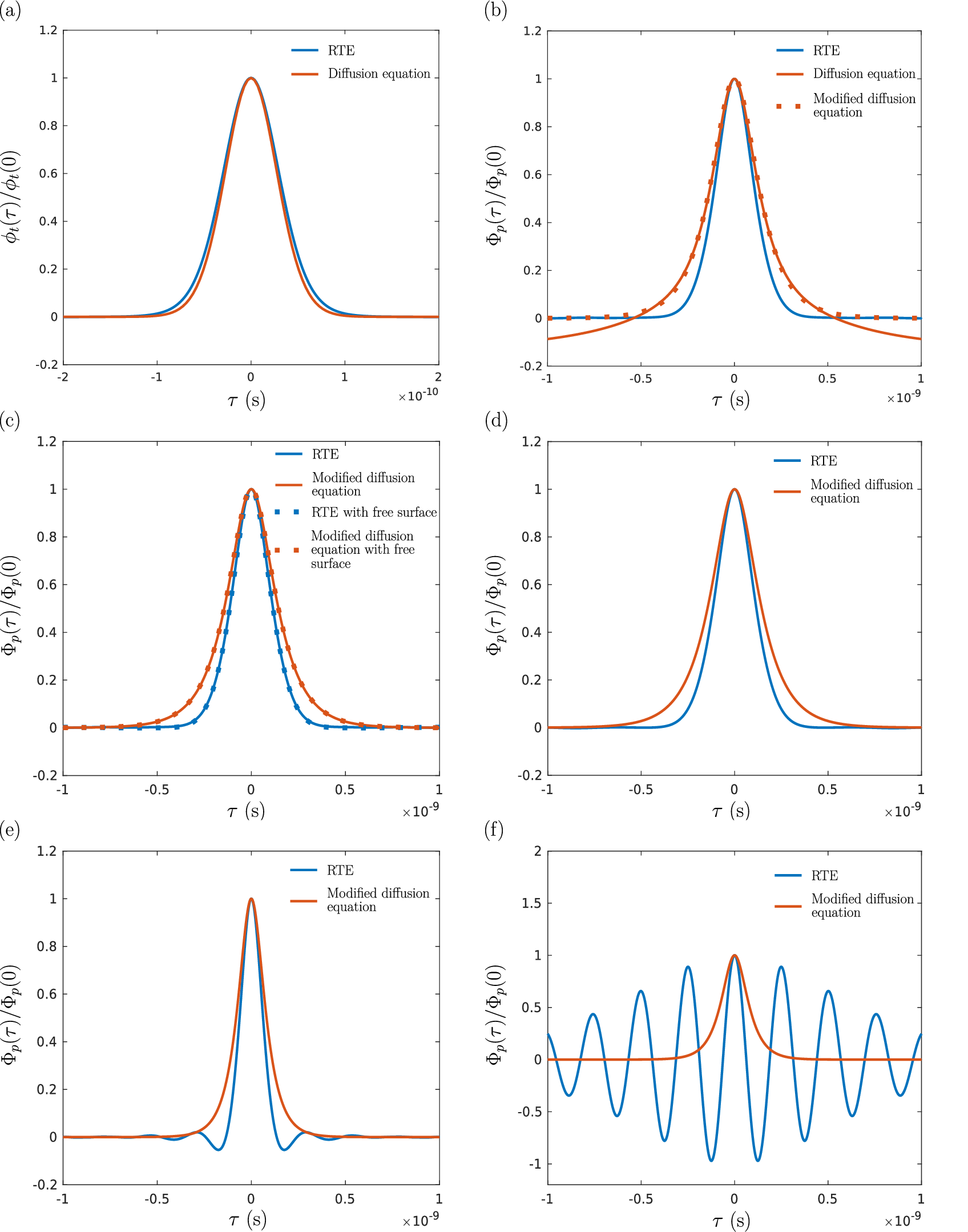}
      \caption{Flux for different situations as a function of the correlation time $\tau$. The
      RTE calculations are in blue and the diffusion equation results are in red.
      (a) Total transmission for the statistical homogeneous medium with isotropic velocity
      distribution and open boundaries for both interfaces.
      (b) Same as (a) but for the flux collected by the probe in reflection. The computation
      using the modified effective absorption term in the diffusion equation is represented by a
      dotted line.
      (c) Same as (b) still with open boundaries (solid lines) or taking into account the free
      surface (dotted lines).
      (d) Same as (c) but with the free surface and a statistically homogeneous lognormal
      particle size distribution and an inhomogeneous particle number density.
      (e) Same as (d) but with an inhomogeneous isotropic velocity distribution.
      (f) Same as (e) but with an inhomogeneous anisotropic velocity distribution.}
      \label{planche_tau}
   \end{figure*}

\begin{figure*}[!ht]
      \centering
      \includegraphics[width=0.80\linewidth]{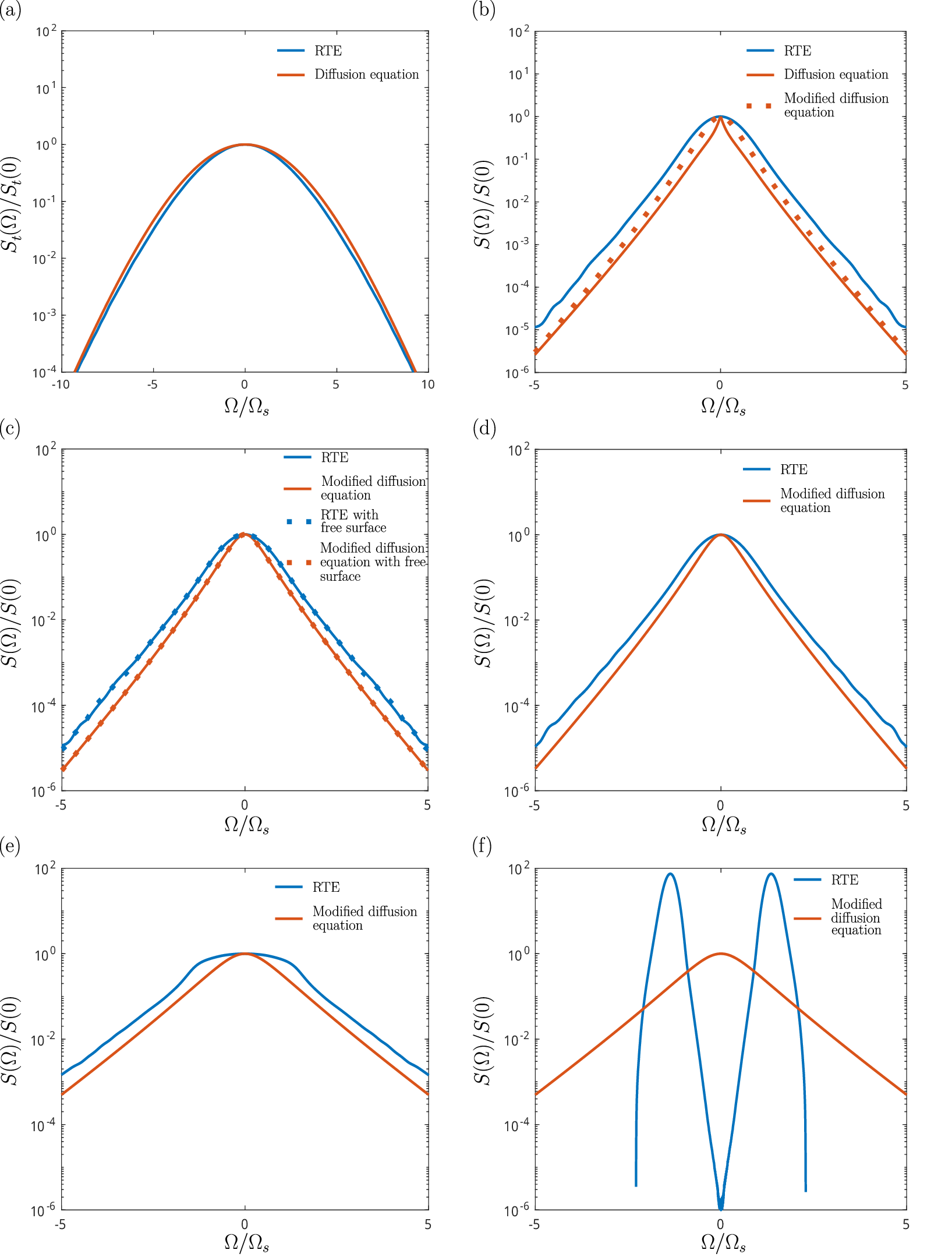}
      \caption{Spectrogram for different situations as a function of the normalized frequency
      $\Omega/\Omega_s$ where $\Omega_s=2 k_r v_s$. The RTE calculations are in blue and the
      diffusion equation results are in red.
      (a) Total transmission for the statistical homogeneous medium with isotropic velocity
      distribution and open boundaries for both interfaces.
      (b) Same as (a) but for the flux collected by the probe in reflection. The computation
      using the modified effective absorption term in the diffusion equation is represented by a
      dotted line.
      (c) Same as (b) still with open boundaries (solid lines) or taking into account the free
      surface (dotted lines).
      (d) Same as (c) but with the free surface and a statistically homogeneous lognormal
      particle size distribution and an inhomogeneous particle number density.
      (e) Same as (d) but with an inhomogeneous isotropic velocity distribution.
      (f) Same as (e) but with an inhomogeneous anisotropic velocity distribution.}
      \label{planche_omega}
   \end{figure*}

   We first focus on the transmitted flux $\phi_t(\tau)$, since the diffusion equation is known
   to be more accurate in transmission. In the diffusion approximation, its expression is given
   by
   \begin{equation}
      \phi_t(\tau)=-\frac{v_E}{2} u_d(z_l,\tau)+I_b^-(z_l).
   \end{equation}
   In practice, the ballistic component is negligible since the optical thickness is large.
   $\phi_t(\tau)$ is plotted in Fig.~\ref{planche_tau}\,(a) and the corresponding spectrogram
   \begin{equation}
      S_t(\Omega)\propto \int \phi_t(\tau)(\cos\Omega\tau)\ud \tau
   \end{equation}
   is plotted in Fig.~\ref{planche_omega}\,(a) with the set of parameters $a_0=\SI{1}{\micro
   m}$, $M_s=\SI{20}{\milli g /\centi m^2}$, $\sigma_v=\SI{1000}{m/s}$, $v_s=\SI{2250}{m/s}$,
   $v_m=\SI{4500}{m/s}$ and $n_r=1$. The plots are represented at time $t=\SI{10}{\micro s}$
   such that $L=\SI{2.25}{\centi m}$. This corresponds to optical thicknesses $b_s=45$ and
   $b_t=L/\ell_t=23$. As a rule of thumb, we assume that $b_t>10$ is needed for the diffusion
   approximation to be valid, which is largely satisfied here. Additionally, we have $g=0.46$.
   As observed in Figs.~\ref{planche_tau}\,(a)~and~\ref{planche_omega}\,(a), a good agreement is
   obtained between the RTE numerical computation and the diffusion model both in the time
   domain and in the frequency domain (spectrogram).

   These results in a simple configuration and in transmission confirm in particular that the
   effective absorption term in the diffusion model correctly renders the decorrelation due to
   motion of the scatterers. They also confirm that it is not necessary to retain a $\tau$
   dependency in the other parameters of the diffusion equation.
 
   We now consider a situation in which the flux is collected in reflection by the probe
   $\phi_p$, and the results are plotted in Figs.~\ref{planche_tau}\,(b) and
   \ref{planche_omega}\,(b). We clearly see some important deviations especially at long
   correlation times and low frequencies. This discrepancy is due to a known limitation of the
   diffusion equation, that it is much less accurate in
   reflection~\cite{pierrat_influence_2008}. Indeed, in this geometry, short paths with few
   scattering events can have a strong impact and are not well accounted for in the diffusion
   approximation. More precisely, the diffusion approximation overestimates the weight of short
   paths and a solution for improvement is to reduce their weight. Since short paths decorrelate
   less than long paths, their influence is mostly visible at long correlation times $\tau$. We
   therefore choose to simply use a second-order Taylor expansion of the effective absorption
   term $1/\tilde{\ell}_a(\tau)$ in terms of $\tau$, in order to increase the effect of the
   effective absorption term at large $\tau$, without affecting the short correlation time
   behavior. To proceed, we use the expression
   \begin{equation}\label{l_a_eff_explicit_dl}
      \frac{1}{\tilde{\ell}_a(\tau)}
      \approx\frac{1}{\ell_a}+\frac{1}{\ell_s}\sigma_v^2k_r^2\tau^2(1-g).
   \end{equation}
   In practice, single backscattering events happening exactly at $z=z_r$ have a null effective
   path length and are therefore not dealt with by the previous correction. To mitigate this
   artefact, the source term in Eq.~(\ref{diffusion_equation_correlation_source_term}) is
   replaced by a Dirac source term at $z_s$ satisfying
   \begin{equation}\label{source_dirac}
      \int_{z_s}^{z_{0,r}}\frac{\ud z}{\ell_t(z)}=1
   \end{equation}
   such that
   \begin{equation}
     S(z,\tau)=I_0\delta\left(z-z_s\right).
   \end{equation}
 
   The results obtained with the diffusion equation modified with the effective absorption term
   in Eq.~(\ref{l_a_eff_explicit_dl}) and the modified source term in Eq.~(\ref{source_dirac})
   are also plotted in Figs.~\ref{planche_tau}\,(b) and \ref{planche_omega}\,(b). Clearly, a
   better agreement is obtained and we will use this improved diffusion model in the following.

   \subsection{Free-surface boundary condition}\label{ejecta_fs}
 
   Having checked the validity of the diffusion model in an open geometry, we can now include
   the free surface by making use of the boundary condition given in Eq.~(\ref{fs}). With this
   boundary condition, it is interesting to compute the reflected flux and compare the result
   with that obtained in the previous configuration with an open boundary and with the same set
   of parameters. Results are presented in Figs.~\ref{planche_tau}\,(c)
   and~\ref{planche_omega}\,(c). We clearly see that the free surface does not affect the
   reflected flux. This is a consequence of the large optical thickness of the ejecta considered
   here. Indeed, few photons propagate over the entire medium and reach the free surface. The
   presence of the free surface has negligible impact on the results.
 
   \subsection{Size distribution and inhomogeneity in the number density}\label{ejecta_a_rho}
 
   We now focus on the influence of the particle size distribution and of the inhomogeneity in
   the particle number density of the ejecta. We start by considering a truncated lognormal size
   distribution for the particle sizes given by
   \begin{equation}\label{lognormal}
      h(a)=
      \begin{cases}
         \displaystyle
         \frac{K}{a\sigma\sqrt{2\pi}}\exp\left[-\frac{\ln^2\left(a/a_0\right)}{2\sigma^2}\right], & \text{if}\,a\in\left[a_\text{min},a_\text{max}\right]\\
         0,&\text{otherwise}
      \end{cases},
   \end{equation}
   where $K$ is a normalization constant given by
   \begin{equation}
      \displaystyle
      K=2\left\{ \text{erf}\left[\frac{\ln\left(a_\text{max}/a_0\right)}{\sigma\sqrt{2}}\right]
      -\text{erf}\left[\frac{\ln\left(a_\text{min}/a_0\right)}{\sigma\sqrt{2}}\right]\right\}^{-1},
   \end{equation}
   with $\text{erf}$ the Gauss error function. Thus, the average particle mass is given by
   $m_p=(4/3)\pi \rho_\text{Sn}\int h(a)a^3 \ud a$ where the integral over $a$ is computed
   numerically. 
 
   We also consider an inhomogeneous particle number density. This inhomogeneity in real ejecta
   is caused by the velocity difference between the slow particles at the back and the fast ones
   at the front, slow particles being more numerous. It is common to describe this inhomogeneity
   using mass-velocity distribution $M(v)$ taking the form of an exponential
   distribution~\cite{monfared_experimental_2014}. Since we are considering a single shock in
   vacuum, we have $v=z/t$ for any particle. Therefore, $M(v)$ does depend on $z/t$ directly and
   reads
   \begin{equation}\label{mv}
      M\left(\frac{z}{t}\right)=
      \begin{cases}
         \displaystyle
         M_s\exp\left(-\beta\frac{z}{v_s t}\right),&\text{if}\,z \in [z_l,z_r]\\
         0,&\text{otherwise}
      \end{cases},
   \end{equation}
   where $M_s$ is the surface mass, the parameter $\beta$ giving the slope of the distribution.
   The particle number density then takes the usual form
   \begin{equation}
      \rho(z)=\frac{\beta M(z/t)}{m_p v_s t}.
   \end{equation}
   Note that $M(v)$ is used at this stage only to derive the expression of the inhomogeneous
   particle number density and does not affect the velocity distribution.  Accounting for both
   the particle size distribution and the inhomogeneity in particle density, the effective
   absorption term takes the form
   \begin{equation}
      \frac{1}{\tilde{\ell}_a(z,\tau)}
      =\frac{1}{\ell_a(z)}+\frac{1}{\ell_s(z)}\sigma_v^2k_r^2\tau^2(1-g).
   \end{equation}

   The results are plotted in Figs.~\ref{planche_tau}\,(d) and~\ref{planche_omega}\,(d) for the
   same parameters as in the previous simulations and $\sigma=0.5$, $a_0=\SI{0.66}{\micro m}$,
   $a_\text{min}=\SI{0.1}{\micro m}$, $a_\text{max}=\SI{2}{\micro m}$, $M_s=\SI{20}{\milli
   g/\centi m^2}$ and $\beta=10$. With these parameters, we solve for the optical thicknesses
   $b_s=42$ and $b_t=\int 1/\ell_t(z)\ud z=23$ and for the anisotropy factor $g=0.47$.

   We observe that the diffusion equation still gives accurate results compared to the RTE. In
   particular, position-dependent parameters in the diffusion model does not break its validity,
   providing that the optical thickness remains large enough for the diffusion approximation
   itself to be valid.
 
   \subsection{Fixed velocity modulus}\label{fixed_modulus}

   In a single shock ejecta, at a given position, the particle velocity is known precisely. To
   improve the description of the ejecta and include this property, we shift from a homogeneous
   isotropic Gaussian velocity distribution to an inhomogeneous isotropic velocity distribution
   with fixed modulus. In terms of statistical distributions for the particle size and velocity,
   this means that we now take
   \begin{align}
      f(z,a,\bm{v}) & = h(a)j\left(z,\bm{v}\right),
   \\
      j(z,\bm{v}) & = \frac{1}{4\pi v_p^2(z)}\delta\left[|\bm{v}|-v_p(z)\right],
   \\
      v_p(z) & = v_s+\frac{z-v_s t}{L}(v_m-v_s).
   \end{align}
   In this setting, the effective absorption mean free path becomes
   \begin{multline}\label{l_a_eff_modulus_z}
      \frac{1}{\tilde{\ell}_a(z,\tau)}
      = \frac{1}{\ell_e(z)}-2\pi\rho(z)
      \int h(a) \sigma_s(a,\omega_0) p(a,\omega_0,\mu)
   \\\times
      \sinc\left[k_r\sqrt{2(1-\mu)}v_p(z)\tau\right]
      \ud a \ud\mu.
   \end{multline}
   Performing the Taylor expansion near $\tau=0$, this expression simplifies into
   \begin{equation}
      \frac{1}{\tilde{\ell}_a(z,\tau)}
      =\frac{1}{\ell_a(z)}+\frac{1}{\ell_s(z)}\frac{v_p^2(z)}{3}k_r^2\tau^2(1-g).
   \end{equation}
 
   The results are plotted in Figs.~\ref{planche_tau}\,(e) and \ref{planche_omega}\,(e) for the
   same parameters as in the previous simulations. We clearly observe a discrepancy between the
   full RTE calculation and the diffusion approximation. In particular, oscillations are visible
   for the RTE model in the $\tau$-domain. These oscillations correspond to the single
   scattering contribution to $\Phi_p$ simply given by $\sinc(\Omega_s \tau)$ in the simplified
   case $v_m=v_s$ (i.e., all particles have the same velocity). Since single scattering events
   are not well captured by the diffusion approximation, these oscillations are not visible in
   the diffusion model. This leads also to the appearance of a break in slope around
   $\Omega=4k_rv_s$ for the spectrogram.

   \subsection{Anisotropic velocity}

   The last property to be included in order to model a realistic ejecta is the anisotropy in
   the particle velocity. Indeed, for a planar shock the velocity must be along $\bm{u}_z$, the
   direction of ejection. This last condition is captured by a velocity distribution of the form
   \begin{equation}
      j(\bm{v},z)=\delta\left[\bm{v}-v_p(z)\bm{u}_z\right].
   \end{equation}
   In these conditions, the effective absorption reads
   \begin{align}\label{l_a_eff_ejecta}
      \frac{1}{\tilde{\ell}_a(z,\tau)}
      ={} & \frac{1}{\ell_e(z)}-2\pi\rho(z)
      \int h(a) \sigma_s(a,\omega_0) p(a,\omega_0,\mu)
   \\ & \times
      \sinc\left[k_r\sqrt{2(1-\mu)}v_p(z)\tau\right]
      \ud a \ud\mu
   \end{align}
   which gives the following Taylor expansion around $\tau=0$
   \begin{equation}
      \frac{1}{\tilde{\ell}_a(z,\tau)}
      =\frac{1}{\ell_a(z)}+\frac{1}{\ell_s(z)}\frac{v_p^2(z)}{3}k_r^2\tau^2(1-g).
   \end{equation}
   We note that the angular velocity distribution has no effect on the effective absorption term
   since an integration is performed over all propagation directions $\bm{u}$ and
   $\bm{u}^\prime$ in Eq.~(\ref{l_a_eff}). In other words, the diffusion model is not sensitive
   to the velocity anisotropy.

   The results are plotted in Figs.~\ref{planche_tau}\,(f) and \ref{planche_omega}\,(f). While
   the diffusive model gives the same results, the RTE gives very different variations compared
   to the previous case. The reason is simple: the diffusion approximation is unable to take
   proper account of the angular distribution of velocities. However, the latter has a
   significant weighting in the low-order scattering events, which in turn have a significant
   weighting in the reflection spectrogram. The diffusive model is thus unable to predict the
   correct behavior of the spectrogram in situations in which the anisotropy in the velocity
   distribution is substantial.

   \section{Conclusion}
 
   In summary, we have derived a model that describes light transport in dynamic media in the
   diffusive regime (\ie, where the transport optical thickness is large). We have shown how a
   model based on the diffusion approximation and accounting for the specificities of an ejecta
   can be derived. This model gives valuable results compared to the RTE in many scenarios
   ranging from a statistically homogeneous and isotropic medium to an ejecta like
   configuration, both in terms of field-field time correlation function and spectrograms. At
   last, we have seen that the model fails to recover the results from the RTE for an isotropic
   velocity distribution with fixed modulus or even worse when the anisotropy of the ejecta
   velocity distribution cannot be neglected. This is a typical bias of the diffusion
   approximation which assumes isotropic properties of the medium and light propagation. For
   standard planar shocks, the RTE therefore remains the reference model. 
   
   Nonetheless, while we have centered this work around planar shocks, the diffusion model
   showed to be very useful in scenarios where the ejecta has a quasi-spherical symmetry. This
   geometry applies for instance to the remarkably interesting work of Saunders et \textit{al.}
   in Ref.~\onlinecite{saunders_experimental_2021} where the interaction of two microjets
   produces a non-planar ejecta. Additionally, a significant advantage of the diffusion model is its
   low computational cost compared to the RTE. This makes it a potentially useful tool for the
   analysis of PDV spectrograms in practical applications.

   Finally, we believe that the idea of developing a simpler transport theory for light
   propagation in ejecta remains relevant and potentially useful in a broad range of practical
   applications. Approaches based on Delta-Eddington phase functions or a Fokker-Plank form of
   the RTE~\cite{gonzalez-rodriguez_comparison_2009,carminati_principles_2021} are possibles
   lines to follow in future works.
 
   \begin{acknowledgments}
      This work has received support under the program ``Investissements d'Avenir'' launched by
      the French Government. 
   \end{acknowledgments}
 
   \section*{Data Availability Statement}
   The data that support the findings of this study are available from the corresponding author
   upon reasonable request. 

%

\end{document}